\documentclass[aps,prl,reprint,groupedaddress]{revtex4-1}
\usepackage{amsmath,amssymb}	
\usepackage{bm}		
\usepackage{ulem}
\usepackage{lineno}			
\usepackage{color}					
\usepackage{graphicx}				
\usepackage{dcolumn}	
\usepackage{textcomp}
\begin{document}
	\newcommand{\By}{$\times$}
	\newcommand{\SqrtBy}[2]{$\sqrt{#1}$\kern0.2ex$\times$\kern-0.2ex$\sqrt{#2}$}
	\newcommand{\Degree}{^\circ }
	\newcommand{\DegreeC}{$^\circ$C }
	\newcommand{\Ohmcm}{$\Omega\cdot$cm}
	\newcommand{\test}{[}
	\newcommand{\kB}{k_\textrm{B}}
	\newcommand{\us}{u_\textrm{s}}
	\newcommand{\ub}{u_\textrm{b}}
	\newcommand{\Ts}{T_\textrm{s}}
	\newcommand{\Tb}{T_\textrm{b}}
	\newcommand{\fin}{f_\textrm{i}}
	\newcommand{\fs}{f_\textrm{s}}
	\newcommand{\Ds}{D_\textrm{s}}
	\newcommand{\Db}{D_\textrm{b}}
	\newcommand{\ds}{d_\textrm{s}}
	\newcommand{\db}{d_\textrm{b}}
	\title{Measurement of the temperature dependence of dwell time and spin relaxation probability of Rb atoms on paraffin surfaces using a beam-scattering method}
	
		\author{Kanta Asakawa$^1$}
		\email{asakawa@go.tuat.ac.jp}
	\author{Yutaro Tanaka$^1$}
	\author{Kenta Uemura$^1$}
	\author{Norihiro Matsuzaka$^2$}
	\author{Kunihiro Nishikawa$^1$}
	\author{Yuki Oguma$^1$}
	\author{Hiroaki Usui$^2$}
	\author{Atsushi Hatakeyama$^1$}
	\email{hatakeya@cc.tuat.ac.jp}
	\affiliation{$^1$Department of Applied Physics, Tokyo University of Agriculture and Technology, Koganei, Tokyo 184-8588, Japan}
	\affiliation{$^2$Department of Organic and Polymer Materials Chemistry, Tokyo University of Agriculture and Technology, Koganei, Tokyo 184-8588, Japan}
	\date{\today}
	\linenumbers
	\begin{abstract}
		The scattering of Rb atoms on an anti-relaxation coating was studied. No significant change in the spin relaxation probability of Rb atoms by single scattering from a tetracontane surface was observed by cooling the film from 305 to 123 K. The mean surface dwell time was estimated using a time-resolved method.
		Delay-time spectra, from which mean surface dwell times can be estimated, were measured at 305, 153, and 123 K, with a time window of $9.3\times 10^{-5}$ s.
		The increase in mean surface dwell time with cooling from 305 to 123 K was smaller than $4.4\times 10^{-6}$ s, which is significantly smaller than the value expected from the mean dwell time at room temperature measured using the Larmor frequency shift. These results can be explained by assuming a small number of scattering components, with a mean surface dwell time at least three orders of magnitude longer than the majority component.
	\end{abstract}	
	
	
	\maketitle
	
	\section{Introduction}
	Anti-relaxation coatings are used to reduce the spin relaxation of alkali metal atoms resulting from wall collisions in the alkali-metal vapor cells of  atomic clocks \cite{risley1980dependence,robinson1982narrow,frueholz1983use} and atomic magnetometers \cite{budker1998nonlinear,balabas2006magnetometry,wasilewski2010quantum}.  Paraffin\cite{robinson1958preservation,bouchiat}, octadecyltrichlorosilane (OTS)\cite{seltzer2007synchronous,zhao2008method}, and polydimethylsiloxane (PDMS)\cite{atutov2016study} are representative materials for anti-relaxation coatings. It has been reported that paraffin-coated surfaces can support $10^4$ spin-preserving collisions for Rb atoms \cite{bouchiat}. The performance of an anti-relaxation coating depends on the surface dwell time, as well as the strength of the interaction between alkali metal spins and the surface. The mean dwell time $\tau_\textrm{s}$ can be described by the Arrhenius formula:
	\begin{equation}
		\tau_\textrm{s}=\tau_0\exp \left(\frac{E_{\textrm{des}}}{\kB  \Ts}\right),\label{arrhenius}
	\end{equation}
	where $\tau_0$ is the pre-exponential factor, $E_{\textrm{des}}$ is the desorption energy, $\kB $ is Boltzmann constant, and $\Ts$ is the temperature of the surface. In the case of Rb atoms on tetracontane (C\mbox{\scriptsize 40}H\mbox{\scriptsize 82}), which is a representative type of paraffin, the experimentally obtained desorption energy is 0.06
	eV\cite{rahman1987rb,budker2005microwave}. By adopting a commonly used assumption, i.e., that the pre-exponential factor is $1\times10^{-12}$ s, which is the typical period of thermal vibration of atoms, we can roughly estimate $\tau_\textrm{s}$ as $1\times 10^{-11}$ s at 300 K. However, using the Larmor frequency shift, Ulanski \textit{ et al.} reported a mean dwell time of $(1.8\pm0.2) \times 10^{-6}$ s for Rb atoms on paraffin coatings at 345 K \cite{ulanski2011measurement}. The reason for the large difference in the mean dwell time calculated from the desorption energy compared to that measured by experiments is still unclear \cite{atutov2015accurate}. One possibility is that the assumption $\tau_0\simeq 1\times10^{-12}$ s is incorrect. By substituting $E_\textrm{des}=0.06$ eV, $\Ts=345$ K, and $\tau_\textrm{s}=1.8\times 10^{-6}$ s into equation (\ref{arrhenius}) and regarding $\tau_0$ as a variable, we obtain $\tau_0=2.4\times 10^{-7}$ s. However, this is two orders of magnitude larger than the pre-exponential factor $\tau_0=2.2\times 10^{-9}$ s of $^{87}$Rb atoms on Pyrex glass surfaces coated with OTS, estimated from the temperature dependence of the mean dwell time \cite{zhao2009method} and is five orders of magnitude larger than the typical period of thermal vibration. Therefore, this issue requires further investigation.  
	
	Equation (\ref{arrhenius}) shows that the mean dwell time increases with cooling, which makes it easier to measure dwell times using time-resolved methods. If we assume that $\tau_0=2.4\times 10^{-7}$ s  (which is obtained by substituting $E_\textrm{des}=0.06$ eV \cite{rahman1987rb,budker2005microwave}, $\tau_\textrm{s}=1.8 \times 10^{-6}$ s, and $\Ts=345$ K \cite{ulanski2011measurement}) is correct, then it can be seen from Eq. (\ref{arrhenius}) that $\tau_\textrm{s}$ will increase by $6.7\times 10^{-5}$ s with cooling of a sample from 305 to 123 K, which is sufficient time to detect using time-resolved methods. The spin relaxation probability with surface scattering is also expected to increase at low temperatures due to an increased dwell time. 
	
	In this study, we investigated the temperature dependence of the spin relaxation probability and dwell time of Rb atoms on tetracontane coatings. A beam-scattering method and X-ray photoelectron spectroscopy (XPS) were employed to analyze the surface dwell time and surface chemical composition. Using an atomic beam and optical hyperfine pumping, the dwell time can be measured more directly compared to the methods used in earlier studies \cite{ulanski2011measurement,zhao2009method}. The results show that the increase in mean dwell time (averaged over the majority of scattered atoms) with cooling from 305 to 123 K was shorter than $4.4\times 10^{-6}$ s, which is significantly shorter than the value of $6.7\times 10^{-5}$ s expected from the previously reported desorption energy $E_\textrm{des}=0.06$ eV \cite{rahman1987rb,budker2005microwave} and the mean dwell time $\tau_\textrm{s}=1.8 \times 10^{-6}$ s at 345 K \cite{ulanski2011measurement}. This indicates the existence of minor scattering components with dwell times at least three orders of magnitude larger than that of the major component.

	\section{Experimental}
	\begin{figure}
		\centering
		\includegraphics[width=0.9\linewidth]{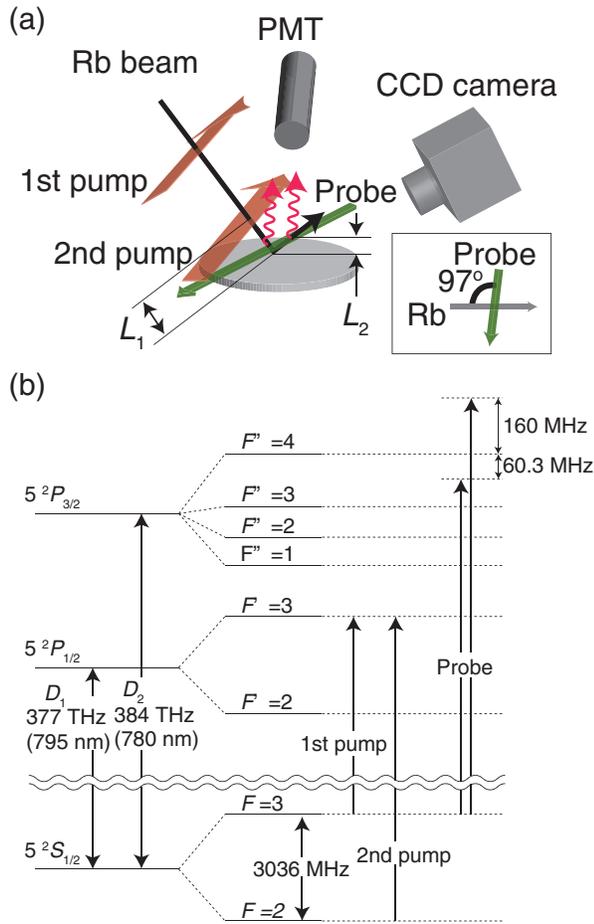}
		\caption{(a) Schematic illustration of the experimental setup and (b) the energy-level diagram of $^{85}$Rb.}
		\label{fig:configuration}
	\end{figure}
	\begin{figure}
		\centering
		\includegraphics[width=0.9\linewidth]{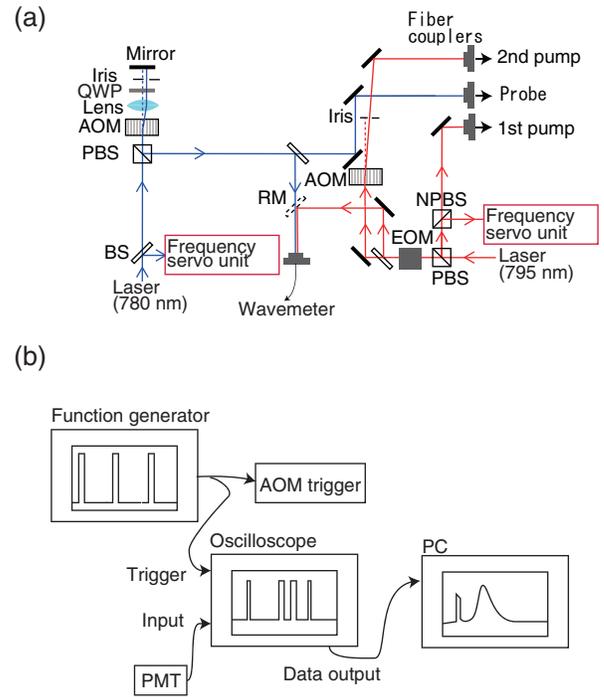}
		\caption{(a) Schematic diagram of the optical system for generating pump and probe light. AOM, EOM, BS, PBS, NPBS, RM, and QWP denote the acousto-optic modulator, electro-optic modulator, beam splitter, polarizing beam splitter, non-polarizing beam splitter, removable mirror, and quarter-wave plate, respectively. (b) Schematic diagram of the signal-processing system used for the delay-time measurements. }
		\label{fig:optics}
	\end{figure}
	
	Figure \ref{fig:configuration}(a) shows the experimental setup. A tetracontane-coated quartz substrate was mounted in an ultra-high vacuum (UHV) chamber with a base pressure lower than $3\times 10^{-7}$ Pa. The tetracontane film was deposited on the substrate in another high-vacuum chamber, the base pressure of which was $1.0\times10^{-5}$ Pa, by evaporating  tetracontane at 513 K for 10 min. The thickness and average roughness, \textit{Ra}, of the film were measured to be $0.93\pm0.17$ \textmu m and 50 nm, respectively, using atomic force microscopy. The Rb beam was generated using a multi-channel effusive atomic beam source. The full width at half maximum of the atomic beam at the position of the sample was estimated to be 8.2 mm from a fluorescence image taken with a charge-coupled device (CCD) camera.
	
	The two pump light beams were directed perpendicular to the Rb beam. The pump and probe frequencies are shown in Fig. \ref{fig:configuration}(b). The optical system used to generate the pump and probe light is illustrated in Fig. \ref{fig:optics}(a). The frequency of the first pump light was tuned to the $F=3 \rightarrow F'=3$  transition frequency of the $^{85}$Rb $D_1$ transition line using polarization spectroscopy \cite{wieman1976doppler,harris2006polarization}, where $F$ and $F'$ are the total angular momentum of atoms in the $5^2S_{1/2}$ and $5^2P_{1/2}$ states, respectively. The second pump light, the frequency of which was tuned to the $F=2 \rightarrow F'=3$  transition frequency of the $D_1$ line, was generated by blue-detuning the first pump light by 3,036 MHz \cite{arimondo1977experimental,schultz2008measurement} using an electro-optic modulator (EOM) and an acousto-optic modulator (AOM), and was pulsed to $5\times 10^{-6}$ s by the AOM. 
	
	The probe light was used to selectively excite Rb atoms in the $F=3$ state to the $F''=4$ state through the $D_2$ transition, resulting in fluorescence. Here, $F''$ is the total angular momentum of the atoms in the $5^2P_{3/2}$ state. The fluorescence was detected by a CCD camera and a photomultiplier tube (PMT), which were equipped with interference filters that were designed to transmit only the probe light and fluorescence. The frequency of the probe light could be tuned, which enabled selective excitation of the incident or scattered atoms. For probing scattered atoms, the probe light was blue-detuned from the $F=3 \rightarrow F''=4$ transition frequency of the $D_2 $ transition line by 160 MHz, such that it did not excite incident atoms. With this frequency, the probe light excites Rb atoms whose velocity component along the probe light is 125$\pm$5 m/s; these are abundant among scattered atoms \cite{sekiguchi2018scattering} but negligible among incident atoms. Here, the natural line width of the Rb $D_2$ transition line (6.06 MHz) \cite{volz1996precision} was used to calculate the uncertainty.
	When probing the incident atoms, the direction of the probe light beam was the same as that used to probe the scattered atoms; however, the frequency was red-detuned by 60.3 MHz from the $F=3\rightarrow F''=4$ transition frequency. With this frequency, the probe light excites atoms with a velocity component along the probe light of $-47 \pm 5$ m/s. Because atoms with this velocity component are found among both incident and scattered atoms, the fluorescence intensity of the incident atoms was estimated by subtracting the contribution of scattered atoms from the measured fluorescence intensity, as discussed below.
	
	Mean dwell-time estimates were based on time-of-flight (TOF) measurements obtained using the pump and probe light. The incident Rb atoms were first pumped to the $F=2$ state by the first pump light and subsequently irradiated with the second pump light, which was pulsed; atoms that were irradiated with the second pump light were momentarily pumped to the $F=3$ state. The incident atoms pumped to the $F=3$ state by the second pump light enhanced the fluorescence induced by the probe light when they reached it. The delay time in the fluorescence enhancement induced by irradiation by the second pump light is the sum of the TOF of the Rb atoms (from the second pump light to the probe light via the film surface) and the surface dwell time. The probe-light-induced fluorescence was detected by the PMT. The signal from the PMT was processed using the system shown in Fig. \ref{fig:optics}(b). Delay-time spectra were acquired by accumulating the time intervals between the irradiation of the second probe light and the detection of fluorescence by PMT. The delay-time distribution can be treated as the distribution of the sum of the TOF and the dwell time only when the hyperfine relaxation by a single collision is negligibly small. When the hyperfine relaxation by a single collision is significant, hyperfine polarization of incident atoms is lost over the surface dwell time. In this case, a large percentage of the scattered atoms experience spin relaxation while on the surface; as such, they do not contribute to the delay-time spectra. Therefore, the probability of spin relaxation resulting from surface scattering must be estimated prior to measurement of the dwell time.
	
	To evaluate the spin relaxation resulting from a single collision, we used the first pump light and the probe light. The first pump light polarizes incident Rb atoms in the beam to the $F=2$ state. Given that the probe-light-induced fluorescence of the incident and scattered atoms reflects the number of atoms in the $F=3$ state, the fluorescence intensity decreases when the first pump light is introduced. The population fraction $\fin $ of the $F=2$ state of the incident atoms pumped by the first pump light can be written as 
	\begin{eqnarray}
		\fin &=&\frac{N_2}{N_2+N_3}\\
		&=&1-\frac{N_3}{N}	,
	\end{eqnarray}
	where $N_2$ and $N_3$ are the numbers of atoms in the $F=2$ and $F=3$ states in the incident atoms, respectively, and $N=N_2+N_3$. Because the fluorescence intensity is proportional to the number of atoms in the $F=3$ state,
	\begin{eqnarray}
		N_3&=&CI_{\textrm{i,p}},
	\end{eqnarray}
	where $C$ is a constant and $I_{\textrm{i,p}}$ is the intensity of the fluorescence of incident atoms induced by the first pump light. $I_\textrm{i,p}$ can be obtained by
	\begin{eqnarray}
		I_{\textrm{i,p}}&=&i_{-60.3\textrm{ MHz,p}} \nonumber\\
		&&-i_{160\textrm{ MHz,p}}\times \frac{M(-47 \textrm{ m/s},\Ts)}{M (125 \textrm{ m/s},\Ts)}\label{Eq.I_ip}
	\end{eqnarray}
	where $i_{\delta,\textrm{p}}$ is the fluorescence intensity measured with the first pump light introduced with the probe light blue-detuned by $\delta$ from the $F=3 \rightarrow F''=4$ transition frequency, and $M(v,\Ts)$ is the Maxwell distribution given by
	\begin{eqnarray}
		M(v,\Ts)=\sqrt{\frac{2m}{\pi \kB \Ts}}\exp\left(-\frac{mv^2}{2\kB \Ts}\right),
	\end{eqnarray}
	where $m$ is the mass of an $^{85}$Rb atom. The first term in Eq. (\ref{Eq.I_ip}) includes contributions from both incident and scattered atoms. To subtract the latter, the second term is introduced. The second term is the fluorescence intensity of the scattered atoms excited by the probe light red-detuned by 60.3 MHz estimated from the fluorescence intensity measured with the probe light blue-detuned by 160 MHz (based on the fact that the velocity distribution of the scattered atoms can be expressed as a Maxwell distribution \cite{sekiguchi2018scattering}).
	
	Because  $F=2$ and $F=3$ states have five- and seven-fold degeneracy, $N_3$ and $N$ can be written as
	\begin{eqnarray}
		N&=&\frac{12}{7}CI_{\textrm{i,np}}.
	\end{eqnarray}
	Here, $I_{\textrm{i,np}}$ is the intensity of the fluorescence of incident atoms without the first pump light, which can be obtained by
	\begin{eqnarray}
		I_{\textrm{i,np}}&=&i_{-60.3\textrm{ MHz,np}}\nonumber \\
		&&-i_{160\textrm{ MHz,np}}\times \frac{M(47 \textrm{ m/s},\Ts)}{M (125 \textrm{ m/s},\Ts)},
	\end{eqnarray}
	where $i_{\delta,np}$ is the intensity of fluorescence measured with the probe light blue-detuned by $\delta$ with the first pump light blocked. Therefore, we can experimentally determine $\fin $ as
	\begin{eqnarray}
		\fin  =1-\frac{7}{12}\frac{I_{\textrm{i,p}}}{I_{\textrm{i,np}}}.
	\end{eqnarray}
	Similarly, the population fraction $\fs$ of the $F=2$ state of the scattered atoms is written as
	\begin{eqnarray}
		\fs =1-\frac{7}{12}\cdot\frac{I_\textrm{{s,p}}}{I_\textrm{{s,np}}},
	\end{eqnarray}
	where $I_{\textrm{s,p}}$ and $I_{\textrm{s,np}}$ are the intensities of the fluorescence of the scattered atoms measured with the probe light blue-detuned by 160 MHz with and without the pump light, respectively. The fluorescence of the incident atoms is negligible when the probe laser is blue-detuned by 160 MHz, which was confirmed by the fluorescence intensity being lower than the detection limit when the sample was removed from the atomic beam position. By comparing $\fin $ and $\fs $, the proportion of atoms whose total angular momentum is changed by scattering at the surface can be estimated.
	
	The delay-time and spin-relaxation measurements were conducted using different samples prepared by the same procedure, to minimize  aggregation of Rb atoms, which may contaminate the surface. The aggregation of Rb on the surface resulting from atomic Rb beam irradiation was checked by XPS using Al k$\alpha$ radiation with a photon energy of 1486.6 eV.
	
	\section{Results and discussion}
	
	\subsection{X-ray photoelectron spectroscopy}
	
\begin{figure}
		\centering
		\includegraphics[width=0.9\linewidth]{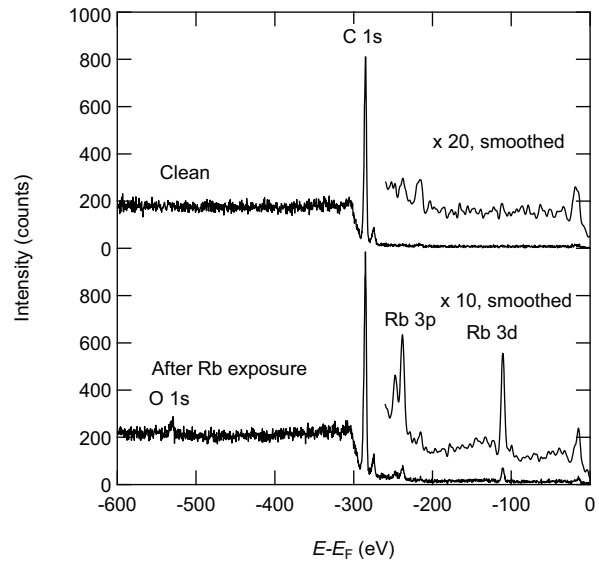}
		\caption{X-ray photoelectron spectroscopy spectra of the clean sample and the sample exposed to an Rb beam at 153 K for 2 h and 123 K for 2 h. }
		\label{fig:xps}
	\end{figure}

	Aggregation of Rb atoms on the surface was investigated by XPS. The temperature at which aggregation was noticeable was approximately 123 K; this varied slightly among samples. Figure \ref{fig:xps} shows the XPS spectra of the as-prepared sample and the spectra taken after exposing the sample to an atomic Rb beam with a flux of $10^{11}$--$10^{12}$ s$^{-1}$ at 153 K for 2 h and 123 K for 2 h. The spectrum of the as-prepared sample displayed a strong peak at $E-E_\textrm{F}=-285$ eV, which was assigned to the C $1s$ state. After exposure to the Rb beam at low temperature, new peaks appeared at $E-E_\textrm{F}=-531$,  $-246$, $-238$, and $-109$ eV; these were assigned to the O $2p$, Rb $3p_{1/2}$, Rb $3p_{3/2}$, and Rb $3d$ states, respectively. The existence of the O $2p$ peak indicates that some of the adsorbed Rb atoms had become oxidized by the residual O$_2$ or H$_2$O in the UHV chamber.
	
	\begin{figure}
		\centering
		\includegraphics[width=0.9\linewidth]{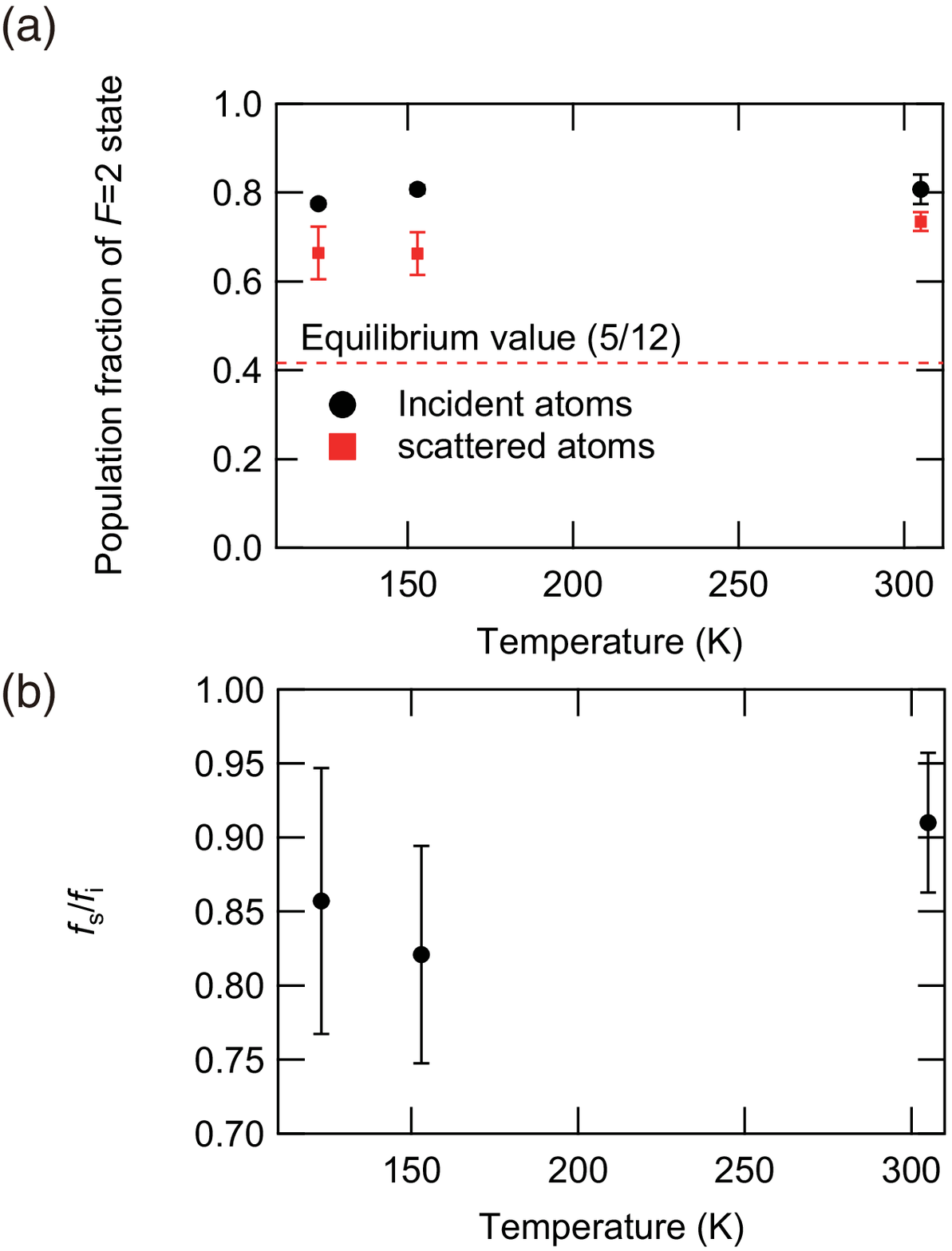}
		\caption{(a)Temperature dependence of the $F=2$ population fraction of the incident and scattered atoms, $f_\textrm{i}$ and $f_\textrm{s}$, respectively and (b) the temperature dependence of the ratio $\fs / \fin $ of the incident and scattered atoms.}
		\label{fig:polarizationscattered}
	\end{figure}
	
	\subsection{Spin relaxation resulting from surface scattering}
	The spin relaxation caused by surface scattering was evaluated at 305, 153, and 123 K. The temperature of the Rb oven of the Rb beam source was set to 393 K. Under these conditions, the flux intensity was estimated to be  $10^{11}$ -- $10^{12}$ atoms per second based on the designed value and the fluorescence induced by the probe light. Below 123 K, the number of scattered atoms was significantly smaller than at above 123 K, indicating the initiation of Rb atom adsorption at around 123 K, which is consistent with the XPS results.
	Figure \ref{fig:polarizationscattered}(a) shows the population fractions of the $F=2$ state for the incident and scattered atoms. Uncertainties in the population fractions were estimated by repeating the measurements three to five times.  Because the spin relaxation induced by a single scattering process is negligibly small at room temperature \cite{bouchiat},  the difference between $f_\textrm{s}$ and $f_\textrm{i}$ at 305 K is attributed to the incident atoms away from the beam center, as opposed to relaxation due to scattering. When measuring the population fraction of the $F=2$ state of incident atoms, atoms that pass the edge of the atomic beam are difficult to pump or detect due to the large deviation in velocity direction with respect to the major component of the incident atoms, which the pump and probe light frequencies are tuned to excite. However, when the probe light is blue-detuned to detect only the scattered atoms, atoms that were not pumped can be detected, as surface scattering changes the direction of the translational movement of atoms. The temperature dependence of the ratio of the population fraction of the incident and scattered atoms $f_\textrm{s}/f_\textrm{i}$ is shown in Fig. \ref{fig:polarizationscattered}(b); no increase in spin relaxation probability induced by cooling was observed above 123 K within the experimental error. Given that Rb atoms experience $10^4$ collisions before their spins relax \cite{bouchiat} in paraffin-coated cells at room temperature, the low spin relaxation probability at 123 K means that the mean dwell time at 123 K is smaller than $10^4$ times the mean dwell time at 300 K. Thus, from Eq. (\ref{arrhenius}), $\tau_0 \exp \left(\frac{E_\textrm{des}}{\kB \cdot 123\textrm{ K}}\right)\leq 10^4 \cdot \tau_0 \exp \left(\frac{E_\textrm{des}}{\kB\cdot 305\textrm{ K}}\right)$. By solving this, we obtain $E_\textrm{des}\leq 0.16$ eV, which agrees with $E_\textrm{des}=0.06$ eV from previous reports \cite{budker2005microwave,rahman1987rb}. 
	
	\subsection{Delay-time spectra}
	\begin{figure}
		\centering
		\includegraphics[width=0.9\linewidth]{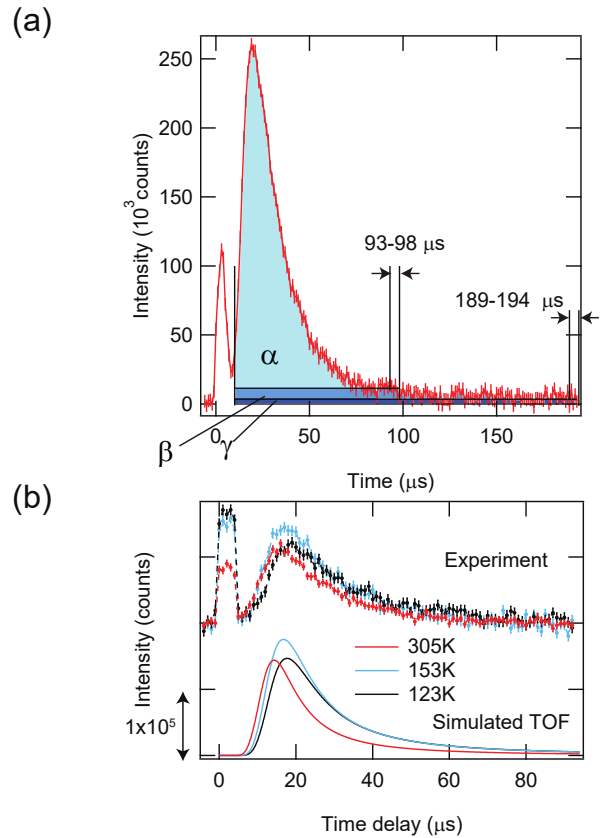}
		\caption{Delay-time spectra (a) at 303 K taken with a high-intensity Rb beam and long repetition period (b) at 305, 153, and 123 K. Dots with error bars represent the experimental data and solid lines represent the simulation results.}
		\label{fig:tof}
	\end{figure}
	Figure \ref{fig:tof}(a) shows the delay-time spectrum at $\Ts=303$ K. The intervals of the second pump light pulses were $2.00\times10^{-4}$ s. During the measurement, spectra with and without the second pump light were acquired by switching the second pump light repeatedly using a shutter. The delay-time spectrum was obtained by subtracting the latter from the former. For this measurement, the temperature of the Rb oven of the beam source was set to 453 K, which was 60 K higher than that used for the low-temperature measurement, to achieve higher signal intensity. As a result, the signal intensity was enhanced by a factor of $\sim 20$. The feature at 0--$1.0\times10^{-5}$ s is attributed to the second pump light, which partially penetrated the interference filter, and the fluorescence of the incident atoms; this is excluded from the intensity integration discussed below. The time origin was defined by the rising edge of the second pump light-derived feature. The feature peaking at around $2\times 10^{-5}$ s was attributed to the enhanced fluorescence of scattered atoms caused by the second pump light. The peak area obtained by integrating the intensity in the region $1.0\times 10^{-5}$--$1.89\times 10^{-4}$ s and subtracting the average level of the region $1.89\times 10^{-4}$--$1.94\times 10^{-4}$ s as the base level, which corresponds to the sum of the area of the regions represented by $\alpha$ and $\beta$ in Fig. \ref{fig:tof} (a), accounted for $91\pm5 \%$ of the total signal intensity. Here, the total signal intensity is the sum of regions $\alpha$, $\beta$, and $\gamma$. If we adopt the average level of the region $9.3\times 10^{-5}$ --$9.8\times 10^{-5}$  s as the base level, the integrated intensity in the region $1.0\times 10^{-5}$--$9.3\times 10^{-5}$ s, which corresponds to the area of region $\alpha$, accounts for $79 \pm2 \%$ of the total intensity. This implies that $79\pm 2 \%$ of the scattered atoms will contribute to the peak intensity in the delay-time spectra, if we regard the sum of the region $t<9.3\times 10^{-5}$ s as the peak intensity and adopt the average level of the region $9.3\times 10^{-5}$ --$9.8\times 10^{-5}$ s as the base level. For the temperature-dependence measurement, we used $1.00\times 10^{-4}$  s as the second pump light interval and subtracted the average of the region near the back edge of the time window from the whole spectra, instead of subtracting the background spectra taken without the second pump light. This dramatically reduced the measurement time, which was essential to prevent Rb aggregation during measurements at low temperatures. 
	
	Figure \ref{fig:tof} (b) shows the delay-time spectra taken at 305, 153, and 123 K. The temperature of the Rb oven of the Rb beam source was set to 393 K. By cooling the tetracontane film from 305 to 123 K, the delay-time spectrum shifted to the longer side and the mean delay time $\tau_\textrm{M}$ increased by $(7.0\pm 3.2)\times 10^{-6}$ s from $(2.54\pm0.21)\times 10^{-5}$  to  $(3.24\pm0.24)\times 10^{-5}$ s. Here,  $\tau_\textrm{M}$ is defined by
	\begin{eqnarray}
		\tau_\textrm{M}=\frac{\sum_{i} t_iI_i}{\sum_{i} I_i},
	\end{eqnarray}
	where $t_i$ and $I_i$ are the delay time and the intensity at the $i$th point, respectively. The uncertainty in	$\tau_\textrm{M}$ originates from the uncertainty in $I_i$ at each point, which is $\sqrt{I_i}$. Data points in the region $6 \times 10^{-6}\leq t_i <9.3\times 10^{-5}$ s were included in the summation. The average of the region $9.5\times 10^{-5} \textrm{ s}\leq t_i< 9.9 \times 10^{-5} \textrm{ s}$ was adopted as the base level.
	Intensities between 0 and $6\times 10^{-6}$ s, which include the peak originating from the second pump light, were not included in the summation. Because the velocity distribution of the scattered beam also depends on the film temperature \cite{sekiguchi2018scattering}, we cannot simply attribute the increase in $\tau_\textrm{M}$ to the increase in mean dwell time. To evaluate the increase in TOF due to the change in velocity distribution, we simulated TOF spectra without taking the dwell time into account.
	
	The simulation considered the TOF from the second pump light to the surface, and from the surface to the probe light. The velocity distribution $\db (v)$ of the incident atoms, and the TOF distribution from the second pump light to the surface $\Db (t,L_1)$, were calculated using equations \cite{sekiguchi2018scattering}
	\begin{eqnarray}
		\db (v)&=&\frac{m^2}{2\kB ^2\Tb ^2}v^3\exp\left(-\frac{mv^2}{2\kB \Tb }\right),\\
		\Db (t,L_1)&=&\db \left(\frac{L_1}{t}\right)\frac{\textrm{d}}{\textrm{dt}}\left(\frac{L_1}{t}\right)\\
		&=& \frac{m^2L_1^4}{2\kB ^2\Tb ^2t^5}\exp \left(-\frac{m}{2\kB  \Tb  }\left(\frac{L_1}{t}\right)^2\right),
	\end{eqnarray}
	where  $\Tb $ is the temperature of the incident beam determined by the temperature of the capillary of the beam source, $v$ is the velocity of atoms, $L_1$ is the distance between the second pump light and the surface along the atomic beam direction, and $t$ is the time. $\Tb $ was 453 K, and $L_1$ was roughly estimated to be $1.8\times 10^{-3}$ m. The distribution $d_\textrm{s}(v_{\perp s})$ of the velocity component of the scattered atoms perpendicular to the surface, which are in thermal equilibrium with the film, and the TOF distribution $\Ds (t,L_2)$ are given by
	\begin{eqnarray}
		\ds (v_{\perp \textrm{s}})&=&\sqrt{\frac{2m}{\pi \kB \Ts }}\exp\left(-\frac{mv_{\perp \textrm{s}}^2}{2\kB T_\textrm{s}}\right),\\
		\Ds (t,L_2)&=&\ds \left(\frac{L_2}{t}\right)\frac{\textrm{d}}{\textrm{dt}}\left(\frac{L_2}{t}\right)\\
		&=&\sqrt{\frac{2m}{\pi \kB \Ts }}\frac{L_2}{t^2}\exp\left(-\frac{m}{2\kB \Ts }\left(\frac{L_2}{t}\right)^2\right),
	\end{eqnarray}
	where $v_{\perp \textrm{s}}$ is the velocity component perpendicular to the film surface, and $L_2$, which was estimated to be $1.42\times 10^{-3}$ m, is the height of the probe  light from the film surface \cite{sekiguchi2018scattering}. The total TOF spectrum $S(t)$ is given by
	\begin{eqnarray}
		S(t)&=&\frac{1}{w}\int_{t-w}^{t}dr\int_{0}^{r}ds\textrm{  }\ub (s,L_1)\nonumber\\
		&&\times \us (r-s,L_\textrm{s}),
	\end{eqnarray}
	where
	\begin{eqnarray}
		\ub (t,L_1)&=&\int_{0}^{\infty}\frac{1}{ \sqrt{2\pi\sigma_1^2} }\exp\left(-\frac{(x-L_1 )^2}{2\sigma_1^2}\right)\nonumber\\
		&\times&\Db(t,x)dx,\\
		\us (t,L_2)&=&\int_{0}^{\infty}\frac{1}{ \sqrt{2\pi\sigma_2^2} }\exp\left(-\frac{(x-L_2 )^2}{2\sigma_2^2}\right)\nonumber\\
		&\times&g_\textrm{s}(t,x)dx,
	\end{eqnarray}
	where $w$ is the duration of the second pump light and $\sigma_1$ and $\sigma_2$ are the $1/\sqrt{e}$ half width of the second pump light and probe light, respectively. $\sigma_1$ and $\sigma_2$ were $2.4\times 10^{-4}$ and $1.7\times10^{-4}$ m, respectively.
	
	The simulation results are indicated by solid lines in Fig. \ref{fig:tof}. The intensities of the simulation results were adjusted to fit the experimental results. The simulated TOF spectra are in good agreement with the experimental results. 
	According to the simulation results at 305 K, the peak area calculated by integrating the spectral intensity between $1.0\times 10^{-5}$ and $9.3\times 10^{-5}$ s and subtracting the average of the region $9.3\times 10^{-5}$--$9.8\times 10^{-5}$ s  accounts for 81\% of the total intensity, which is in good agreement with the experimental results shown in Fig. \ref{fig:tof}(a).
	The simulated mean TOF calculated from the region 0--9.3 $\times 10^{-5}$ s increased by $5.8 \times 10^{-6}$ s from $2.48\times 10^{-5}$ to $3.06\times 10^{-5}$ s with cooling from 305 to 123 K. Here, the average of the region $9.5\times 10^{-5} \textrm{ s}\leq t_i< 9.9 \times 10^{-5} \textrm{ s}$ was adopted as the base level. The experimentally observed shift of $\tau_\textrm{M}$, which is $(7.0\pm3.2) \times 10^{-6}$ s, is the sum of the increase in mean dwell time and mean TOF. Therefore, the increase in mean dwell time induced by cooling can be obtained by subtracting the increase in the simulated mean TOF from the experimentally obtained increase in the mean delay time. Therefore, we can see from the experimental and simulation results that 
	\begin{eqnarray}
		\tau_\textrm{s,$t< 93$ \textmu s}(123\textrm{ K})-\tau_\textrm{s,$t< 93$ \textmu s}(305\textrm{ K})\nonumber\\
		=(1.2\pm3.2)\times 10^{-6}\textrm{ s}, 
	\end{eqnarray}	
	which means  
	\begin{eqnarray}
		\tau_\textrm{s,$t< 93$ \textmu s}(123\textrm{ K})-\tau_\textrm{s,$t< 93$ \textmu s} (305\textrm{ K})\nonumber\\
		\leq 4.4\times 10^{-6}  \textrm{ s.} \label{increaseoftau}
	\end{eqnarray}	
	Here, $\tau_{\textrm{s},t< t_{\textrm{max}}}(\Ts)$ is the mean dwell time of the scattering component, with a delay time of less than $t_{\textrm{max}}$ at a surface temperature of $\Ts$.
	
	\subsection{Discussion}
	
	The discrepancy between our result ($\tau_\textrm{s}(123\textrm{ K})-\tau_\textrm{s}(305\textrm{ K})\leq 4.4\times 10^{-6}$ s) and the value ($\tau_\textrm{s}(123\textrm{ K})-\tau_\textrm{s}(305\textrm{ K})=6.7\times 10^{-5}$ s) obtained by substituting the desorption energy  $E_{\textrm{des}}=0.06$ eV  \cite{budker2005microwave,rahman1987rb} and mean dwell time $\tau_\textrm{s}=1.8\times10^{-6}$ s at $T=345$ K  \cite{ulanski2011measurement} into Eq. (\ref{arrhenius}) may be explained by assuming multiple scattering components with different mean dwell times. Using the method described in this study, the scattering components with dwell times larger than $9.3\times 10^{-5}$ s is not detected. In Ref. \cite{ulanski2011measurement}, on the other hand, the mean dwell time was estimated based on the Larmor frequency shift caused by the interaction with the surface and evanescent pump light.  
	
	The reason for the difference in mean dwell times between the scattering components can be attributed to differences in pre-exponential factors. It has been reported that a certain proportion of the incident atoms penetrate the PDMS film, diffuse into the bulk, and desorb from the surface \cite{atutov2015accurate}, which makes the mean dwell time about a million times larger than that calculated from the desorption energy and film temperature. If the diffusion barrier in the bulk is significantly smaller than the desorption energy, the temperature dependence of the diffusion time can be neglected so that the temperature dependence of the mean dwell time is almost entirely determined by the desorption energy. By assuming that two scattering components with different mean dwell times exist, we can approximate the temperature dependence of the mean dwell time as
	\begin{eqnarray}
		\tau_\textrm{s}=(1-p)\tau_1\exp\left(\frac{E_\textrm{des}}{\kB \Ts }\right)+
		p\tau_2 \exp\left(\frac{E_\textrm{des}}{\kB \Ts }\right),\label{twocomponentarrhenius}
	\end{eqnarray}
	where $p$ is the proportion of scattering events with longer mean dwell times, and $\tau_1$ and $ \tau_2$ are the pre-exponential factors for the scattering events with shorter and longer mean dwell times, respectively.	 We suppose that $\tau_2 \exp\left(\frac{E_\textrm{des}}{\kB \Ts }\right) $ is significantly larger than the time window of $9.3\times10^{-5}$ s and only the component with a shorter mean dwell time, which corresponds to the first term in Eq. (\ref{twocomponentarrhenius}), contributes to the delay-time spectra. By substituting Eq. (\ref{arrhenius}) into our results ($\tau_\textrm{s}(123\textrm{ K})-\tau_\textrm{s}(305\textrm{ K})\leq 4.4\times 10^{-6}$ s), we obtain $0<\tau_1\leq 1.6\times 10^{-8}$ s. From $\tau_2 \exp\left(\frac{E_\textrm{des}}{\kB \Ts }\right)\gg 9.3\times10^{-5}$ s  at $T_\textrm{s}\leq305$ K, we get $\tau_2\gg 9.47 \times 10^{-6}$ s. By substituting $\tau_\textrm{s}=1.8\times10^{-6}$ s at 345 K \cite{ulanski2011measurement} and $E_\textrm{des}=0.06 $ eV \cite{rahman1987rb,budker2005microwave} into Eq. (\ref{twocomponentarrhenius}), $p=\frac{2.4\times 10^{-7} \textrm{s}-\tau_1}{\tau_2-\tau_1}$. From $0<\tau_1$ and $\tau_2\gg 9.47\times 10^{-6}$ s, $p<0.025$, which means that the component with a shorter mean dwell time is the major component. This is consistent with the fact that the observed $79\pm 2$ \% fraction within $9.3\times 10^{-5}$ s is nearly the same as the 81 \% fraction obtained from the simulation without dwell times.
	
	\section{Conclusions}
Scattering	of Rb atoms on tetracontane surfaces was investigated. No significant spin relaxation was observed with a single scattering process down to 123 K. The temperature evolution of delay time  showed that the increase in mean surface dwell time induced by cooling from 305 to 123 K was less than $4.4\times 10^{-6}$ s. Taken together, the results indicate the existence of multiple scattering sites. The pre-exponential factor $\tau_0$ of the minor components is at least three orders of magnitude larger than that of the major component, which means that the mean dwell time of the minor scattering components is at least three orders of magnitude larger than that of the major component. 
	\section{Acknowledgments}
	
	This work was supported by JSPS KAKENHI Grant Number JP17H02933.

\end{document}